\documentclass[%
 reprint,
 amsmath,amssymb,
 aps,
]{revtex4-2}
\usepackage{makecell}
\usepackage{graphicx}
\usepackage{dcolumn}
\usepackage{bm}
\usepackage{hyperref}

\DeclareUnicodeCharacter{2061}{}
\begin{document}

\preprint{APS/123-QED}

\title{Anomalous Bilayer Quantum Hall Effect}

\author{Gurjyot Sethi\textsuperscript{1}}
 \author{D. N. Sheng\textsuperscript{2}}
\author{Feng Liu\textsuperscript{1}}
\affiliation{\textsuperscript{1}Department of Materials Science \& Engineering, University of Utah, Salt Lake City, Utah 84112, USA\\ \textsuperscript{2}Department of Physics \& Astronomy, California State University, Northridge, CA 91330, USA}

\date{\today}

\begin{abstract}
In parallel to the condensed-matter realization of quantum Hall (Chern insulators), quantum spin Hall (topological insulators) and fractional quantum Hall (fractional Chern insulators) effects, we propose that bilayer flat band (FB) lattices with one FB in each layer constitute solid-state analogues of bilayer quantum Hall (BQH) system, leading to anomalous BQH (ABQH) effect, without magnetic field. By exact diagonalization of a bilayer Kagome lattice Hamiltonian, as a prototypical example, we demonstrate the stabilization of excitonic condensate Halperin’s $(1,1,1)$ state at the total filling $v_T=1$ of the two FBs. Furthermore, by tuning the inter-layer tunneling and distance between the Kagome layers at $v_T=2/3$, we show phase transitions among Halperin’s $(3,3,0)$, spin-singlet $(1,1,2)$, and particle-hole conjugate of Laughlin’s $1/3$ states, as previously observed in BQH systems. Our work opens a new direction in the field of FB physics by demonstrating bilayer FB materials as an attractive avenue for realizing exotic ABQH states including non-Abelian anyons.
\end{abstract}

\maketitle


Condensed-matter realization of quantum Hall effect (QHE), first proposed by Haldane \cite{1}, has been one of the most significant scientific breakthroughs that instigated the origin of topological states of matter.  Subsequently, quantum spin Hall effect (QSHE) was proposed in solid-state systems by Kane-Mele \cite{2,3} and Bernevig-Hughes-Zhang \cite{4} further advancing this field with critical real material realizations \cite{4,5}. Of particular interest among the latest developments in this ever-growing field is the interplay between topology and electron interaction, inherent to a topological flat band (FB). 2D lattices hosting Chern FB \cite{6,7} have been shown to stabilize fractional QH (FQH) effect without magnetic field, manifesting a fractional Chern insulator (FCI) as a condensed matter analog of FQH \cite{8,9}, recently realized in twisted bilayer graphene system \cite{10}. At fractional filling of FBs, in analogy to filling Landau levels (LLs), ground state mimics the Laughlin’s FQH state with features like ground-state degeneracy on a torus, spectral flow under flux insertion, and quasihole statistics \cite{8,9}.\par
A distinctive property of emergent excitations of FQH state is that they carry fractional charges \cite{11} following anyonic statistics in 2D \cite{12,13}. Especially, the non-Abelian fractional statistics \cite{14} obeyed by FQH states are attractive owing to their potential application in topological quantum computation \cite{15}. However, the realization of non-Abelian anyons in FCIs is not straightforward and require strong $n$-body interactions ($n>2$) \cite{16,17}, since in a single-component FQH system, non-Abelian anyons are realized by such interactions in partially filled excited LLs ($v_T>1$) rather than the ground-state LL.\par
\begin{figure}[h]%
\centering
\includegraphics[width=0.48\textwidth]{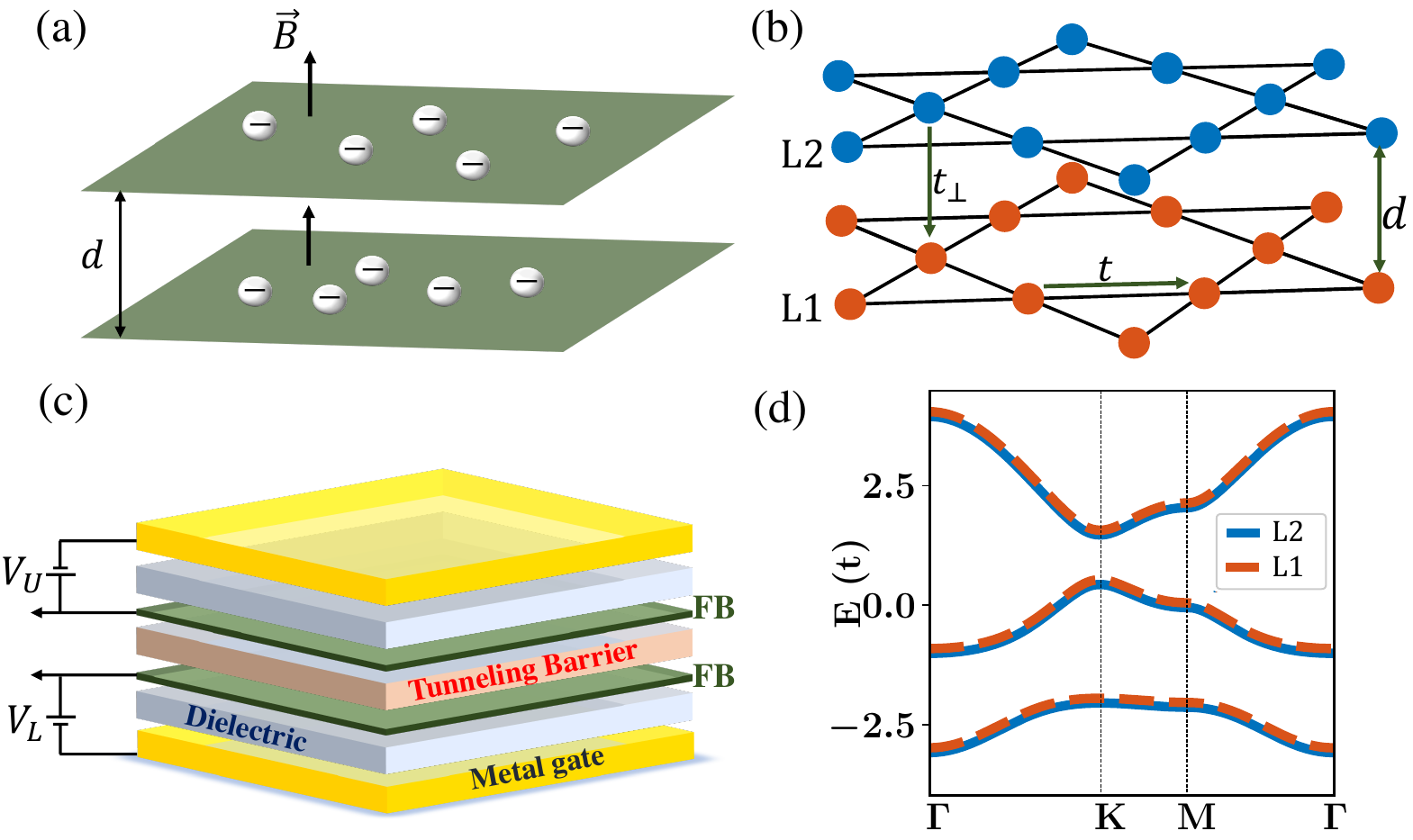}
\caption{(a) Schematic of BQH setup containing two layers of 2DEG separated by a distance $d$ and a tunneling barrier. (b) Schematic of bilayer Kagome lattice with blue and orange colors indicating atoms in layer L1 and L2 respectively. (c) Schematic of possible device configuration that can be used to realize ABQH. V\textsubscript{U(L)} represent gate voltages for upper(lower) layer. FB material is indicated by green layer with a tunneling barrier in between. (d) Single-particle band structure of bilayer Kagome lattice at $\lambda=0.3t$ and $t_{\perp}=0$. The two FBs belonging to individual layers are depicted in blue and orange color respectively.}\label{fig1}
\end{figure}

Beyond single-component FQH, on the other hand, multi-component FQH effects offer a richer and more exotic quantum phase diagram, with extra degrees of freedom allowing for additional tuning parameters \cite{18,19,20,21}. Multi-component FQH systems are described by Halperin's (m,m,n) wavefunctions \cite{18}. In the following, we only use integers (m,m,n) to refer to these states. Bilayer Quantum Hall (BQH) systems, such as GaAs quantum wells (QWs) \cite{22,23,24,25}, are spectacular examples of multi-component FQH effect. Either a wide QW or a double QW can be mapped onto two layers of two-dimensional electron gas (2DEG) separated by a finite distance (Fig.~\ref{fig1}(a)) forming the two-component FQH setup, leading to exotic multi-component QH plateaus \cite{22,23,24,25}. Notably, one of the plateaus manifests at $v_T=1$ ($(1,1,1)$ state) where inter-layer excitonic coherence is observed by counter-flow measurements \cite{26,27,28}. In addition, BQH systems are known to host a plethora of Abelian and non-Abelian FQH phase transitions owing to exceptional tunability of parameters, such as inter-layer distance and tunneling strength, which has generated a lot of theoretical and experimental interest \cite{29,30,31,32}.\par
In this Letter, we propose a largely unexplored research direction, the realization of exotic multi-component BQH states in two topological FBs of same Chern number (to be distinguished from the yin-yang FBs of opposite Chern numbers studied recently \cite{33}) without magnetic field. This is especially appealing since BQH systems can host non-Abelian anyons at $v_T<1$ \cite{21,30,32}. We note that some recent works \cite{34,35,36} have discussed the possibility of stabilizing multi-component bosonic/fermionic FQH states in a single layer topological FB partially filled by spin-up and -down electrons, but these proposals lack the essential inter-component tunability present in BQH systems.\par
In parallel to anomalous QH and FQH effects, we propose the concept of anomalous BQH (ABQH) effect. We provide direct evidence for ABQH states in bilayer Kagome lattices, as a prototypical bilayer FB system. Kagome lattice is one of the most widely studied 2D lattices and material systems hosting topological FBs \cite{7,37,33,39,40}, and excitingly a recent experiment has observed the topological FB in a \textit{monolayer} breathing-Kagome lattice \cite{41}. We note that bilayer-Kagome-lattice materials have also been experimentally synthesized as candidates to host quantum spin liquids \cite{42,43}. In addition, topological FBs are proposed in multi-layer heterostructures \cite{44,45}, as well as in metal organic frameworks \cite{46,47,48,49}. The feasibility in synthesizing multi-layer lattices with individual layer hosting a FB has partly motivated the present study to realize ABQH states without magnetic field. Using exact diagonalization (ED) method, we demonstrate characteristic signatures of ABQH states in a bilayer Kagome lattice. We have calculated topological degeneracy of ground-state manifold on a torus, non-trivial Chern matrix and excitonic off-diagonal long-range order, to prove the stabilization of excitonic condensate states at total filling $v_T=1$ of two FBs, and further illustrate the potential of this lattice in stabilizing exotic tunable ABQH states at $v_T=2/3$, which may enable the realization of non-Abelian anyons.\par
We use a nearest-neighbor (NN) tight-binding model of bilayer Kagome lattice (Fig.~\ref{fig1}(b)) with spin-orbital coupling (SOC) that conserves the out-of-plane spin-component \cite{37}, leading to the kinetic part of total Hamiltonian in the absence of inter-layer tunneling,
\begin{equation}
\begin{split}
    H_{kin}=\sum_{\sigma}[-t&\sum_{<i,j>\alpha}c^{\dagger}_{i\sigma\alpha}c_{j\sigma\alpha}\\&+i\lambda\sum_{<i,j>\alpha\beta}\frac{2}{\sqrt{3}}(r^{1}_{ij}\times r^{2}_{ij}).\tau^z_{\alpha\beta}c^{\dagger}_{i\alpha}c_{j\beta}]\label{eqn1}
\end{split}
\end{equation}
where $t$ is the NN hopping integral, $\lambda$ represents SOC strength, $r^{1,2}_{ij}$ are the two NN unit vectors, $\alpha(\beta)$ are spin indices, $\tau^z$ is the Pauli 's matrix and $\sigma(\sigma')$ represents each layer of the bilayer lattice. Electron interactions are described by an extended Hubbard model including NN intra-layer ($V$) and direct inter-layer ($V_{\perp}$) interactions. $V_{\perp}$ depends on the inter-layer distance $d$ as shown in Fig.~\ref{fig1}(b) which can be independently varied, rendering the tunability of inter-component interactions. We also include tunneling between the layers ($t_{\perp}$) in the interaction part of the Hamiltonian. This is closely related to the experimental observation of excitonic condensate in bilayer graphene where a tunneling barrier is introduced between the two layers to extend the lifetime of electron-hole pairs \cite{50,51}. The strength of $t_{\perp}$ can be varied by changing the tunneling-barrier material in between the layers for a given inter-layer distance. A schematic of possible device is shown in Fig.~\ref{fig1}(c). Hence, the interaction part of Hamiltonian in our model is given as,
\begin{equation}
\begin{split}
    H_{int}=\sum_{\sigma,{\sigma}'}[V\sum_{<r,r'>_{intra}}n_{r\sigma}n_{r'{\sigma}}&+V_{\perp}\sum_{r}n_{r\sigma}n_{r\sigma'}\\&+t_{\perp}\sum_{r} c^{\dagger}_{r\sigma}c_{r\sigma'}]\label{eqn2}
\end{split}
\end{equation}
where $n_{r\sigma}=c^{\dagger}_{r\sigma}c_{r\sigma}$ is the electron density operator. The spin indices are suppressed. Throughout this work we set $V=2t$, which is smaller than the single-particle gap separating the FBs (Fig.~\ref{fig1}(b)), while a distance dependence of $V_{\perp}=1.5t/d$ is assumed.\par
Single-particle band structure is shown in Fig.~\ref{fig1}(d). With $t_{\perp}=0$, the two FBs have the same Chern number and belong to the individual layers of the bilayer Kagome lattice, which exactly match the two LLs in each layer of conventional BQH system. In our model, we use a slightly high SOC ($\lambda=0.3t$) to isolate the bottom FBs from other bands. This procedure is widely used in realizing FCI where a high flatness ratio of FB is desirable \cite{37}. It is worth mentioning that such conditional parameters can be realized using Floquet band engineering in real materials \cite{40,52}. To study the effect of interactions, we exactly diagonalize the full Hamiltonian ($H=H_{kin}+H_{int}$) in reciprocal space with interactions projected to the two FBs of finite lattice containing total number of $N_s=6\times N_x\times N_y$ sites. The total filling factor is given by $v_T=N_e/(N_x\times N_y)$, where $N_e$ is the number of electrons in the system. Under periodic boundary conditions (PBC), we implement translational symmetry and diagonalize the Hamiltonian in each momentum sector $q=(2\pi k_x/N_x ,2\pi k_y/N_y)$ with $k_x$ and $k_y$ being the integers, labelled as $(k_x, k_y)$. One can also assign a pseudospin notation to each FB and define $S_z=(N_{\uparrow}-N_{\downarrow})/2$, where $N_{\uparrow(\downarrow)}$ is the number of electrons in FB belonging to upper (lower) layer. $S_z$ is a good quantum number when $t_{\perp}=0$. Detailed methodology of ED for multiple bands is given in our previous work \cite{53} [also see supplementary material (SM) Sec. A].\par
\begin{figure}[h]%
\centering
\includegraphics[width=0.48\textwidth]{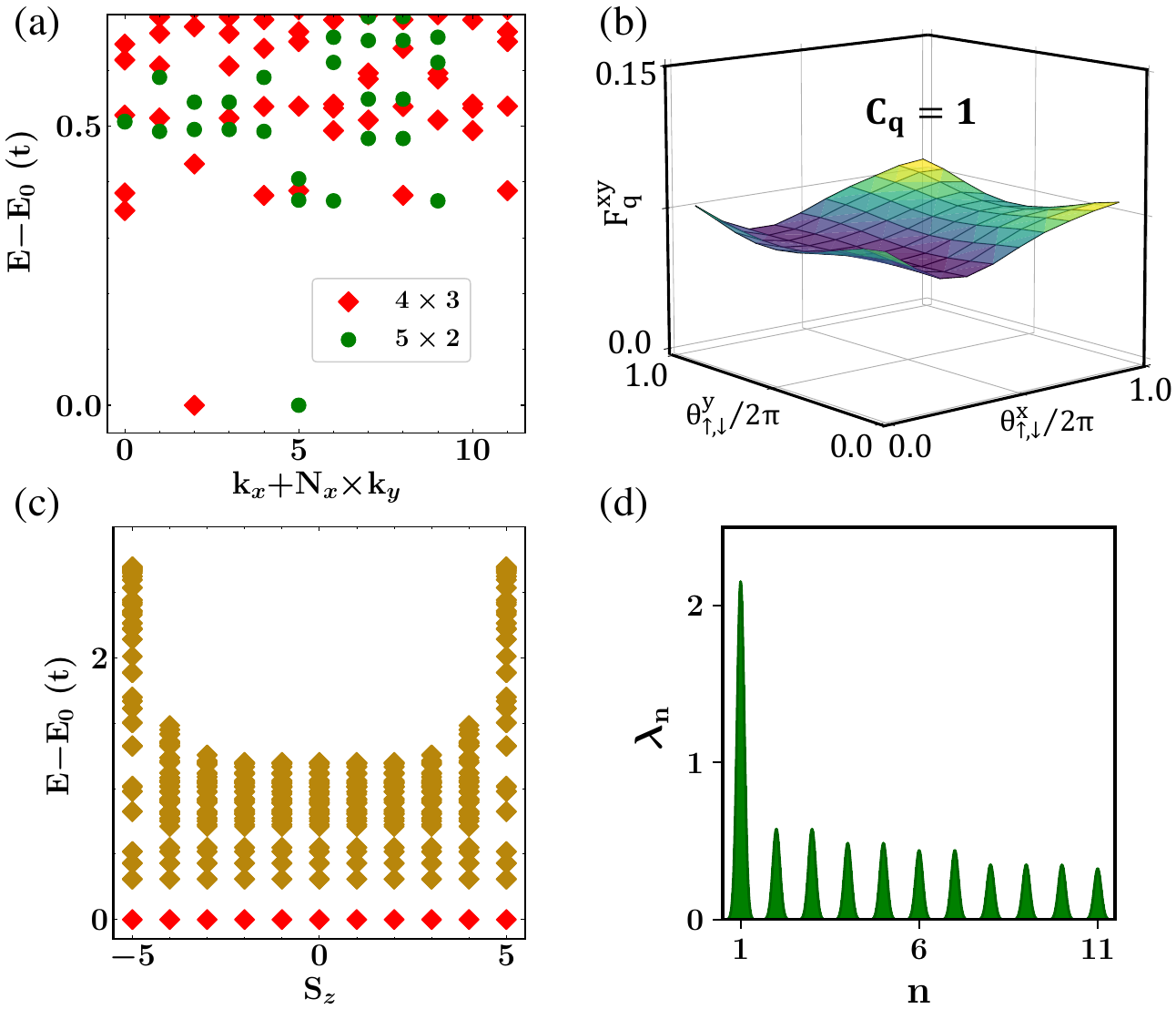}
\caption{(a) Momentum-resolved spectra of $H$ at $d=0.3a$, $t_{\perp}=0$, and $S_z=0$ for two system sizes, $N_s=6\times 5 \times 2$ and $N_s=6 \times 4 \times 3$ denoted by green circles and red diamonds respectively. (b) Berry curvature for total charge Chern number indicated by $C_q$. (c) Energy spectra of $H$ at $d\rightarrow0$ and $t_{\perp}=0$ with varying $S_z$ for $N_s=6\times 4 \times 3$ system size. Red (yellow) diamonds denote the ground (excited) states. (d) Eigenvalues of $\rho^{(2)}$ for (b) plotted in descending order.}\label{fig2}
\end{figure}

First, we study the $v_T=1$ case.  In a BQH setup with negligible $t_{\perp}$, $v_T=1$ plateau manifests excitonic condensate with the wavefunction described by $(1,1,1)$ state featuring a non-degenerate ground state on two tori \cite{36}. In Fig.~\ref{fig2}(a) we plot the energy spectra for $S_z=0$, or the case of balanced layers with $N_{\uparrow}=N_{\downarrow}$, at a finite $d=0.3a$ and negligible $t_{\perp}$. There is clearly presence of a non-degenerate ground state separated from excited states in two finite systems of different sizes, which are shown to confirm the convergence of our results. For all subsequent calculations the system size of $4\times3$ is used. $(1,1,1)$ state on two tori with PBC should belong to the total momentum sector $(\sum_{i=1}^N k_x^i ,\sum_{i=1}^N k_y^i)$, where $N=N_x\times N_y$, implying all the reciprocal points are occupied; for the $4\times 3$ system, it is the $(2,0)$ sector, as shown in Fig.~\ref{fig2}(a), implying a non-trivial topological characteristic of the ground state with a binding of particle-hole between two layers.\par
Beyond the energy spectra, we extract the Chern number matrix $C=\begin{bmatrix}C_{\uparrow \uparrow} & C_{\uparrow \downarrow} \\ C_{\downarrow \uparrow} & C_{\downarrow \downarrow}\end{bmatrix}$ for the two-component FQH system, which contains information for quantized Hall conductance \cite{31,34,35,36}. For Abelian multi-component FQH states, the diagonal and off-diagonal elements of $C$ are related to intra-component and inter-component Hall transports respectively. Each component of $C$ is defined under twisted boundary conditions with $\theta_{\uparrow(\downarrow)}^{x(y)}$ being the twist angle in layer $\uparrow(\downarrow)$ in $x(y)$ direction \cite{34,35,36,55}. With $\theta_{\uparrow}^x=\theta_{\downarrow}^x=\theta^x$, and $\theta_{\uparrow}^y=\theta_{\downarrow}^y=\theta^y$, charge Hall conductance is related to $C_q=\sum_{\sigma,\sigma '}C_{\sigma,\sigma '}$. For the single non-degenerate ground state at $v_T=1$, we numerically calculate Berry curvatures on a $11\times 11$ mesh in the phase space and obtain a topological invariant $C_q=1$ with a well-defined smooth Berry curvature as plotted in Fig.~\ref{fig2}(b), further confirming the non-trivial topology of $(1,1,1)$ state.\par
An important additional feature of $(1,1,1)$ state is the excitonic condensation observed in experiments \cite{27}. In Fig.~\ref{fig2}(c) we show the energy spectra of $H$ for $v_T=1$ with negligible $d$ and $t_{\perp}$ at multiple $S_z$. The degeneracy of states with varying $S_z$ is a signature of spontaneous symmetry breaking, also known as the “which layer” uncertainty for BQH \cite{27} indicating the stabilization of excitonic condensate. At a finite $d$, state with $S_z=0$ becomes energetically stable due to anisotropy in interactions [see SM Sec. B]. We further  illustrate excitonic coherence order by calculating eigenvalues of excitonic reduced two-body density matrix \cite{53}, $\rho^{(2)}(k,k';\overline{k},\overline{k}')=<\Psi \rvert \psi_2^{\dagger}(k)\psi_1(k')\psi_1^{\dagger}(\overline{k}')\psi_2(\overline{k})\rvert \Psi>$, where $\psi_{1(2)}^{\dagger}(k)$ creates an electron in FB belonging to layer $1(2)$ at reciprocal lattice point $k$, and $\rvert \Psi>$ is the many-body ground-state wavefunction. As shown in Fig.~\ref{fig2}(d), there exists one large eigenvalue which concretely illustrates the off-diagonal long-range order of the excitons formed between the occupied states in one FB and empty states in the other \cite{53}, characteristic of a symmetry-broken condensate order parameter.\par
 
\begin{figure}[h]%
\centering
\includegraphics[width=0.48\textwidth]{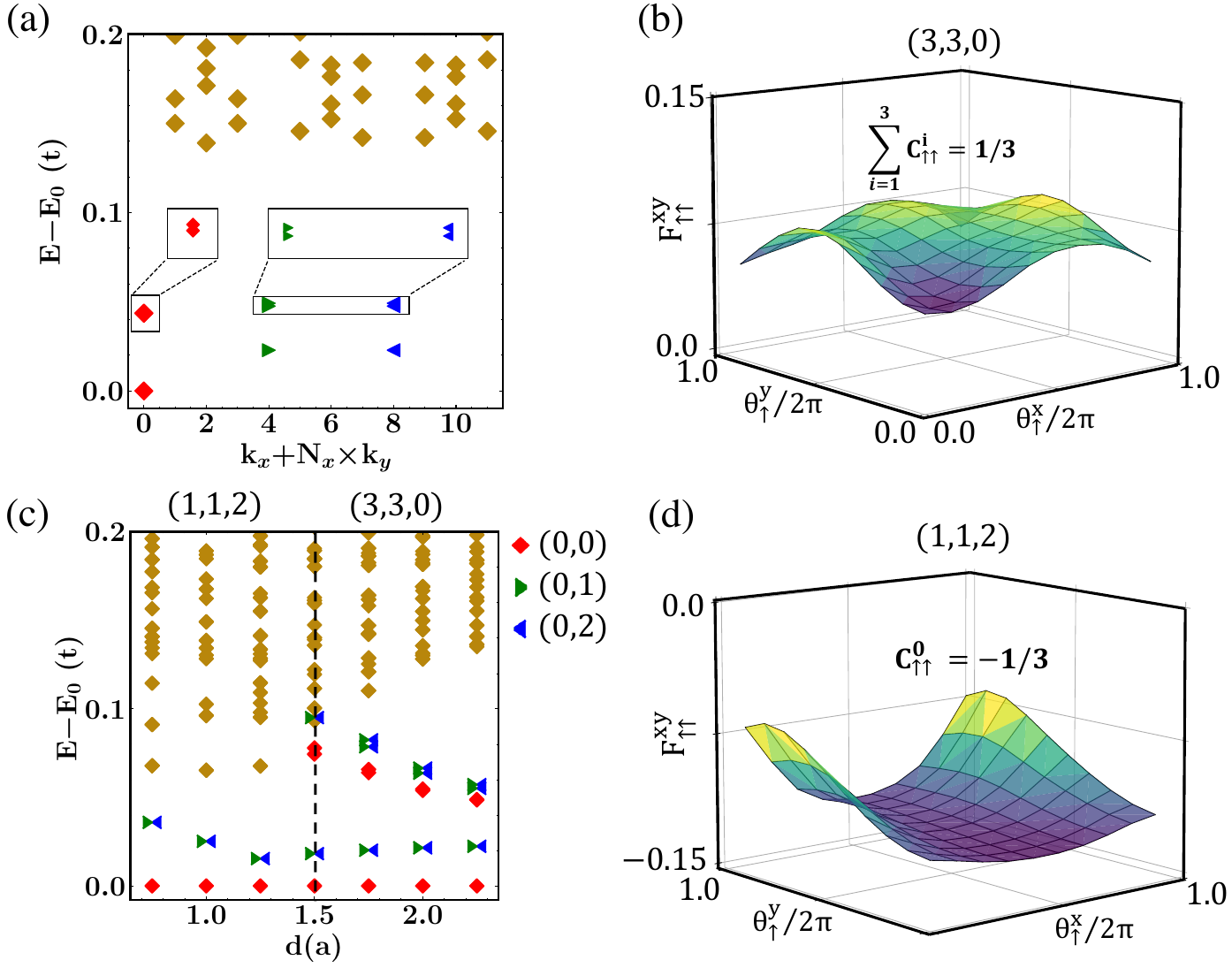}
\caption{(a) Energy spectra of $H$ at $d=2.25a$ and $t_{\perp}=0$ for $N_s=6\times 4 \times 3$ system size. Red diamonds $(0,0)$, blue $(0,1)$ and green $(0,2)$ triangles denote the 9-fold degenerate ground-state manifold while yellow denotes the excited states. Inset shows the zoomed-in view of ground-state manifold (b) Total intra-component Berry curvature ($C_{\uparrow \uparrow}$) of the 3 quasi-degenerate states at $(0,0)$ for system in (a). (c) $d$-driven phase transition from $(1,1,2)$ state to $(3,3,0)$ state. (d) Intra-component Berry curvature ($C_{\uparrow \uparrow}$) for the single state at $(0,0)$ of the 3-fold degenerate ground-state manifold of $(1,1,2)$ calculated at $d=1.0a$.}\label{fig3}
\end{figure}
Next, we focus on filling factor $v_T=2/3$, in order to illustrate the tunability of ABQH states as realized in bilayer Kagome system. At $v_T=2/3$ and a relatively large $d=2.25a$ between the two layers, we plot the energy spectrum for balanced layers $(S_z=0)$, as depicted in Fig.~\ref{fig3}(a). The individual layers are decoupled when $d$ is large and should stabilize $(3,3,0)$ state \cite{32}, which is simply the bilayer Laughlin’s state. We, first, use topological degeneracy of ground-state manifold on a torus to identify the non-trivial character of this state \cite{deg,56,57}. Inter-component correlations in $(m,m,n)$ state, in general, can be studied using $K$-matrix, $K=\begin{bmatrix} m & n \\ n & m\end{bmatrix}$, as formulated within the field theory framework \cite{58,59}. Topological degeneracy of ground state for a system with $v_T=p/q$ on a torus is given by the determinant of $K$-matrix, $\mathrm{det}⁡K=qN^{\prime}$, where $N^{\prime}$ is an integer describing the different translations of the center-of-mass (COM) of different components, while $q$ determines the overall COM degeneracy. In the case of  $(3,3,0)$ state at $v_T=2/3$, and a $4\times 3$ system size, the three degenerate states corresponding to $N^{\prime}=3$ should lie at $(0,0)$ \cite{53}. Taking COM degeneracy into account there should be additional 3-fold degeneracy for each of these states as depicted in Fig.~\ref{fig3}(a).\par
To further illustrate this state to be topologically non-trivial, we calculate the $C$ matrix (which should be $=K^{-1}$ \cite{31}) for the 9 quasi-degenerate ground states of $(3,3,0)$ state. For the three states at $(0,0)$, we found that the intra-component Berry curvature is well-defined and smooth (Fig.~\ref{fig3}(b)), from which we obtain a $C$ matrix element, $\sum_{i=1}^3 C_{\uparrow \uparrow}=1$, while the off-diagonal element of $C$ matrix vanishes $(\sum_{i=1}^3 C_{\uparrow \downarrow}=0)$. Similarly, for all the 9 states in momentum sectors $(0,0)$, $(0,1)$ and $(0,2)$, we obtain $\sum_{i=1}^9 C_{\uparrow \uparrow}=3$ and $\sum_{i=1}^9 C_{\uparrow \downarrow} =0$. The other two elements are related to these two by symmetry $c_{\uparrow} \rightarrow c_{\downarrow}$. This finally gives,
\begin{equation}
    C=\begin{bmatrix} C_{\uparrow \uparrow} & C_{\uparrow \downarrow} \\ C_{\downarrow \uparrow} & C_{\downarrow \downarrow}\end{bmatrix}=\frac{1}{9}\begin{bmatrix} 3 & 0 \\ 0 & 3\end{bmatrix}\label{eqn4}
\end{equation}
which is inverse of the $K$-matrix for $(3,3,0)$ state, proving the topological characteristic of this state. In addition, we calculate the spectral flow of these 9 states under flux insertion, confirming fractional charge Hall conductance (Fig.~\ref{figs3}(a) and (b)).\par
Next, we show a $d$-driven transition from spin-singlet $(1,1,2)$ state at small $d$ to $(3,3,0)$ state at large $d$ in Fig.~\ref{fig3}(c). When $d$ is small, the $v_T=2/3$ ABQH resembles the usual $v_T=2/3$ FQH state with spin, i.e., for small $d$ the pseudospin can be directly mapped onto electronic spin, which stabilizes $(1,1,2)$ state with 3-fold degeneracy on a torus \cite{31}. In Fig.~\ref{fig3}(c) we show a transition from the 3- to 9-fold degenerate ground state, as $d$ increases, with gap closing at $d=1.5a$.  The latter ($d<1.5a$) is identified as $(3,3,0)$ state having the $C$ matrix as shown in Eqn.~\ref{eqn4}; the former ($d >1.5a$) is characterized with $C=\frac{1}{3}\begin{bmatrix}-1 & 2 \\ 2 & -1\end{bmatrix}$, as expected for $(1,1,2)$ state, numerically calculated from the inter- and intra-component Berry curvature (Fig.~\ref{fig3}(d)).\par
\begin{figure}[h]%
\centering
\includegraphics[width=0.48\textwidth]{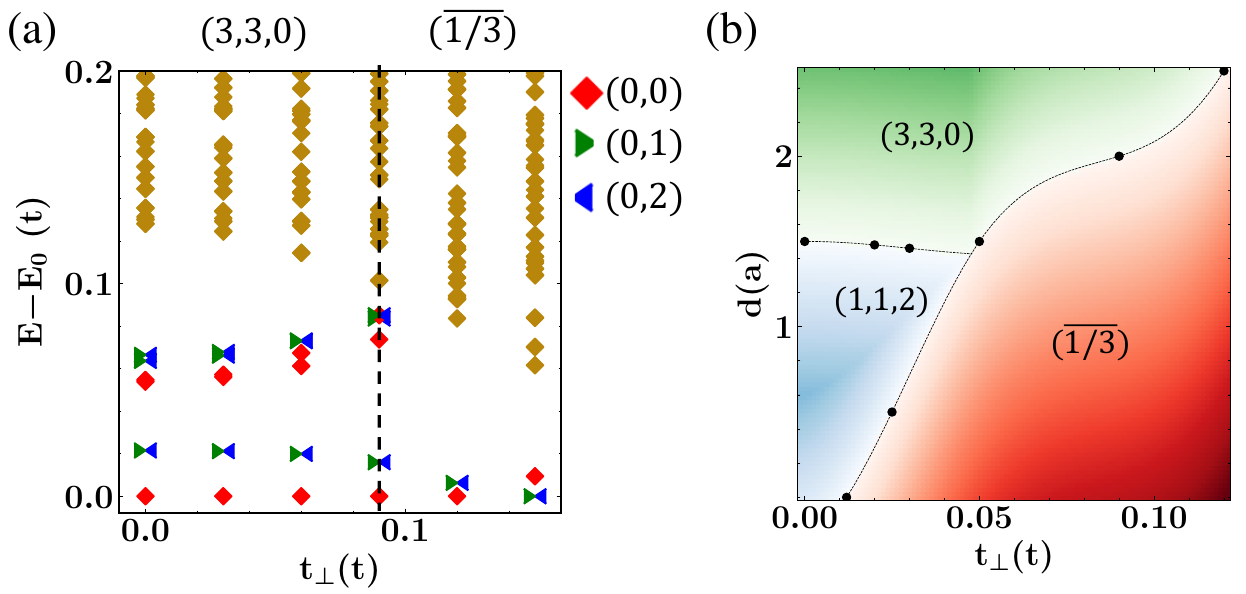}
\caption{(a) $t_{\perp}$-driven phase transition at $v_T=2/3$, $d=2.0a$, $S_z=0$ for $N_s=6\times 4\times 3$ system size. (b) Phase diagram for bilayer Kagome lattice at $v_T=2/3$ in the phase space of $d$ and $t_{\perp}$ between the two Kagome layers}\label{fig4}
\end{figure}
There is also a possibility of $t_{\perp}$-driven transition at $v_T=2/3$ \cite{32}. In Fig.~\ref{fig4}(a) we show a transition from $(3,3,0)$ state to particle-hole conjugate of Laughlin’s $1/3$ state $(\overline{1/3})$ at $d=2.0a$. At low $t_{\perp}$ $(3,3,0)$ state is identified using the $C$ matrix as described earlier. At strong $t_{\perp}$, the bilayer effectively behaves as a single layer due to strong correlations and stabilizes the $\overline{1/3}$  state with 3-fold COM degeneracy at $(0,0)$, $(0,1)$, and $(0,2)$, which is also seen in Fig.~\ref{fig4}(a), similar to BQH system [32]. Hence, in the phase space of $t_{\perp}$ and $d$, we identify three phases: $(3,3,0)$ state at large d and low $t_{\perp}$; $(1,1,2)$ state at small $d$ and low $t_{\perp}$ ; $\overline{1/3}$ state at low $d$ and high $t_{\perp}$, as shown by the phase diagram in Fig.~\ref{fig4}(b). The phase transition boundary from the 3-fold degenerate $(1,1,2)$ to $\overline{1/3}$ state is obtained using gap-closing and reopening points (Fig.~\ref{figs3}(c)). This phase diagram for bilayer FB lattice is in excellent agreement with the one experimentally and theoretically explored for conventioned BQH system \cite{32}.\par
The striking tunability of ABQH states, as illustrated here for bilayer Kagome lattice, could also lead to the material realization of non-Abelian anyonic states without magnetic field. As an example, a non-Abelian inter-layer Pfaffian state with 9-fold ground-state degeneracy is proposed for $v_T=2/3$ BQH when inter-layer interaction becomes slightly attractive \cite{32}. We see a similar 9-fold degeneracy of ground-state manifold in the ABQH state at negative inter-layer interaction strength in our calculations (Fig.~\ref{figs4}(a)), but more concrete evidence requires the analysis of orbital-cut entanglement spectra.  Moreover, at $v_T=1/2$ a $t_{\perp}$ driven transition from $(3,3,1)$ to non-Abelian Moore-Read state is proposed for BQH \cite{30}. We observe also the features of $(3,3,1)$ at $v_T=1/2$ in bilayer Kagome system (Fig.~\ref{figs4}(b)), however, the investigation of the transition to Moore-Read Pfaffian state requires numerical calculations with much larger system size as shown earlier \cite{30}. Further investigation of these exotic non-Abelian states has been left for future work. \par
One important aspect of realizing the exotic ABQH states in bilayer FB lattices is their experimental feasibility. Recently, FCI was experimentally identified in twisted bilayer graphene over hexagonal boron nitride substrate under a weak magnetic field \cite{10}. A potential material candidate to realize ABQH states could be double twisted bilayer graphene separated by a tunneling barrier (Fig.~\ref{fig1}(b)). Moreover, Floquet engineering of isolated FBs have been shown as a promising route towards realizing FCIs in real 2D materials, for example, graphene \cite{52}, and a metal-organic 2D monolayer with Kagome bands \cite{40}. This approach can be generalized to bilayers of such FB materials. One possibility is using Floquet engineering to isolate FBs in the bilayer of superatomic graphene lattice, whose single-layer structure has already been experimentally synthesized using molecular building blocks of triangulene \cite{60} [also see SM Sec. D]. With the present work, therefore, we have demonstrated the role of unique topological properties of FBs in realizing exotic ABQH states and provided a new solid-state platform to study and experimentally realize Abelian/non-Abelian anyonic statistics in multi-component FQH systems.\par
We thank Y.S. Wu for helpful discussions. G.S. and F.L. acknowledge support from US Department of Energy (DOE)-Basic Energy Sciences (Grant No. DE-FG02- 04ER46148). D.N.S. acknowledges support by the DOE-Basic Energy Sciences under Grant No. DE-FG02-06ER46305 for studying topological states. All calculations were done on the CHPC at the University of Utah and NERSC.

\section*{Supplementary Material}
\subsection{Computational Details}
\subsubsection{Many-body calculation setup}
As shown by Eqn.~\ref{eqn1}, kinetic energy part of the many-exciton Hamiltonian is based on a tight-binding (TB) model of bilayer Kagome lattice with two flat bands (FBs). The interaction part of the Hamiltonian is given by Eqn.~\ref{eqn2}. We further project the Hubbard interactions onto the two FBs leading to the projected interaction term,
\begin{align}
&H_{int}^{proj}=\frac{V_n}{N}\sum_{k_i}\delta^{2\pi}_{k_1+k_3-k_2-k_4}\sum_{<x,y>_n}V^{xy}_{k_1,k_2,k_3,k_4}\times\nonumber\\
&\times\sum_{\alpha_i}u^*_{xk_1,\alpha_1}u_{xk_2,\alpha_2}u^*_{yk_3,\alpha_3}u_{yk_4,\alpha_4}c_{\alpha_1k_1}^{\dagger}c_{\alpha_2k_2}c_{\alpha_3k_3}^{\dagger}c_{\alpha_4k_4},\label{eqn5}
\end{align}
where $V^{xy}$ is phase factor acquired by the pair $<x,y>$ of $n^{th}$ NN after projection, {$\alpha_i$} represent all combinations of the two FBs belonging to the two layers of bilayer lattice, $u_{xk,\alpha}$ is the $x$ component of $\alpha$-band Bloch wavefunction calculated at reciprocal point $k$, and $N$ is the finite system size, i.e., the number of allowed reciprocal lattice points. The interaction term includes all the inter- and intra-band interactions from which we neglect the terms that cause Coulomb induced excitations. This is done since the electron density in each layer is fixed by gate voltages (Fig.~\ref{fig1}(c)) The many-body basis states are given by $\prod_{k'}c_{\uparrow,k'}^{\dagger}\prod_k c_{\downarrow,k}^{\dagger}\rvert 0>$, where $\rvert0>$ is defined as the state with empty FBs, and $c_{\uparrow (\downarrow),k}^{\dagger}$ creates an electron at reciprocal lattice point $k$ in the FB belong to upper (lower) layer of the bilayer lattice. We then exactly diagonalize the many-body projected Hamiltonian.
\subsubsection{Exact Diagonalization (ED) method for solving many-body Hamiltonian}
ED method, used for calculating many-body wavefunctions and energies, is known for its computationally expensive nature, both in terms of time and memory \cite{61}. Therefore, it heavily relies on the use of high-performance computational architecture. This method has been previously used for studying mostly fractional Chern insulators \cite{9}, where many-body basis states comprise of the possible ways a fixed number of electrons can partially fill the topological flat band (FB). 

In this work we use the same methodology but extend it to more than one band. Since in the absence of tunneling between the two FBs, our projected Hubbard interaction terms conserve the number of electrons in each band ($S_z$ is conserved) as well as the total electronic momentum ($k_{tot}$), as can be seen from Eqn.~\ref{eqn5}, we use these symmetries to reduce the dimensions of Hilbert space. We solve $H$ for each $S_z$ and $k_{tot}$ sector separately. To illustrate our methodology, here we use a fictitious system of two FBs with finite system size of $3\times1$ which implies there are 3 allowed reciprocal lattice momenta: 0, $2\pi/3$, and $4\pi/3$, and work with $v_T=2/3$, i.e. $N_e=2$. We illustrate below methodology for $S_z=0$ ($N_{\uparrow}=N_{\downarrow}=1$) and $k_{tot}=0$. A typical ED method involves three steps:

\textbf{Basis states formation.} As mentioned above, the basis states in the many-body Hilbert space for our model are given by $\prod_{k'}c_{\uparrow,k'}^{\dagger}\prod_kc_{\downarrow,k}^{\dagger}\rvert0>$. For our fictitious system, there are $C_1^3$  possible ways that 1 electron can occupy 3 allowed reciprocal momenta in each FB. For the block with total electronic momenta = 0, the basis set comprises of 3 states: 
\begin{align}
&c_{\uparrow,0}^{\dagger}c_{\downarrow,0}^{\dagger}\vert0>,\nonumber \\
&c_{\uparrow,4\pi/3}^{\dagger}c_{\downarrow,2\pi/3}^{\dagger}\vert0>,\nonumber \\
&c_{\uparrow,2\pi/3}^{\dagger}c_{\downarrow,4\pi/3}^{\dagger}\vert0>. \nonumber
\end{align}
The ordering of creation and annihilation operators are kept consistent throughout. Our ED code also employs efficient lookup tables for the basis states as their number can reach $\sim10^8$.

\textbf{Hamiltonian matrix element computation.} Once the basis set is created, the next step is to construct the many-body Hamiltonian matrix. This requires operating all the terms in the bands-projected Hubbard interaction (Eqn.~\ref{eqn5}) on each basis state. Parallel implementation and memory-mapped I/O are used to store the upper half of this Hermitian matrix on disk, which can reach $\sim1$ TB in disk space. An example of one of the terms acting on one of the basis states for the fictitious system is 
\begin{align}
(c_{\downarrow,0}^{\dagger}c_{\uparrow,0}^{\dagger}c_{\uparrow,4\pi/3}c_{\downarrow,2\pi/3})c_{\uparrow,4\pi/3}^{\dagger}c_{\downarrow,2\pi/3}^{\dagger}\rvert0>=c_{\uparrow,0}^{\dagger}c_{\downarrow,0}^{\dagger}\rvert0>.\nonumber
\end{align}
Here we have used fermionic commutation relation to be consistent with basis ordering convention. Note that the interaction term used in this example is the inter-FB interaction term and conserves total electronic momentum.

\textbf{Diagonalizing the Hamiltonian.} After the matrix is constructed and stored on disc, we use Lanczos algorithm \cite{62} to find the first few lowest eigenvalues of Hamiltonian. We also compute the wavefunctions and store them for later analysis.

\subsubsection{Twisted boundary conditions, Chern number matrix, \& spectral flow}
Twisted boundary conditions ($\psi(r_{\sigma}+N_{\sigma})=\psi(r_{\sigma}) \mathrm{exp}⁡(i\theta_{\sigma}^{\alpha})$, where $\theta_{\sigma}^{\alpha}$ is the twist angle in $\alpha$ direction for electrons in layer $\sigma$) are implemented by transforming each single-particle momentum $k_{x(y)}\rightarrow k_{x(y)} +\theta/N_{x(y)}$. Many-body Berry curvature is calculated using wavefunctions obtained in the phase space of twist angles for each FB. The Chern number of the many-body ground state in the parameter space $(\theta_{\sigma}^x,\theta_{\sigma '}^y)$ is given by,
\begin{equation}
    C_{\sigma,\sigma '}=\int{\int{\frac{d\theta_{\sigma}^x d\theta^y_{\sigma '}}{2\pi}F_{\sigma,\sigma '}^{xy}}}\label{eqn3}
\end{equation}
where $F_{\sigma,\sigma '}^{xy}=2 \mathrm{Im} (< \frac{\partial \psi}{\partial \theta_{\sigma}^x} \rvert \frac{\partial \psi}{\partial \theta_{\sigma '}^y}>)$ is the Berry curvature \cite{34,35,36}. The twisted boundary conditions can also be used to simulate flux insertion into the system. This allows us to perform spectral flow analysis (see sections II, III, and IV). We use Laughlin’s gauge argument \cite{9}, according to which if one adiabatically inserts $m$ quantum flux into the $1/m$ FQH state, the states should evolve back to their original configuration and there should not be any mixing between the states in ground-state manifold and the excited states throughout flux insertion.

\subsubsection{Degeneracy of many-body ground states on a torus}
Topological degeneracy of ground-state manifold on a torus is a powerful tool that can be used to identify the FQH phases \cite{56,57}. This was shown earlier by Haldane \cite{56} for single component FQH and later expanded by McDonald \cite{57} for multi-component FQH on a torus. For multi-component FQH systems, inter-component correlations in Halperin’s $(m,m,n)$ state can be studied using $K$-matrix, $K=\begin{bmatrix} m & n \\ n & m \end{bmatrix}$, as formulated within the framework of field theory \cite{56,57}. Topological degeneracy of ground state for a system with $v_T=p/q$ on a torus is given by the determinant of $K$-matrix, $\mathrm{det}⁡K=qN^{\prime}$, where $N^{\prime}$ is an integer describing the different translations of the center-of-mass (COM) of different components, while $q$ determines the overall COM degeneracy. Moreover, $N^{\prime}$ can be expressed as the sum of $N_0$, the number of states at $(0,0)$, and $N_B$, the number of states at zone boundary momentum sectors $(0,\overline{N}/2)$, $(\overline{N}/2,0)$, and $(\overline{N}/2,\overline{N}/2)$, where $\overline{N}=N_e/p$. For homogenous liquids, $N^{\prime}=N_0+3N_B$ with two possibilities: if $N_e/(N_i p)$ is even, $N_0=N_B=N^{\prime}/4$, otherwise, $N_0=N^{\prime}$ and $N_B=0$, where $N_i$ is the number of electrons with component $i$. The $q$ COM degenerate points are given by Pauli’s generalized principle \cite{9} - if one state in the ground-state manifold lies in the momentum sector $(k_1,k_2)$, the next state can always be found at $(k_1+N_e,k_2+N_e)$ [modulo $(N_x,N_y)$].
\subsection{Halperin’s (1,1,1) state}
\begin{figure}
\centering
\includegraphics[scale=0.4]{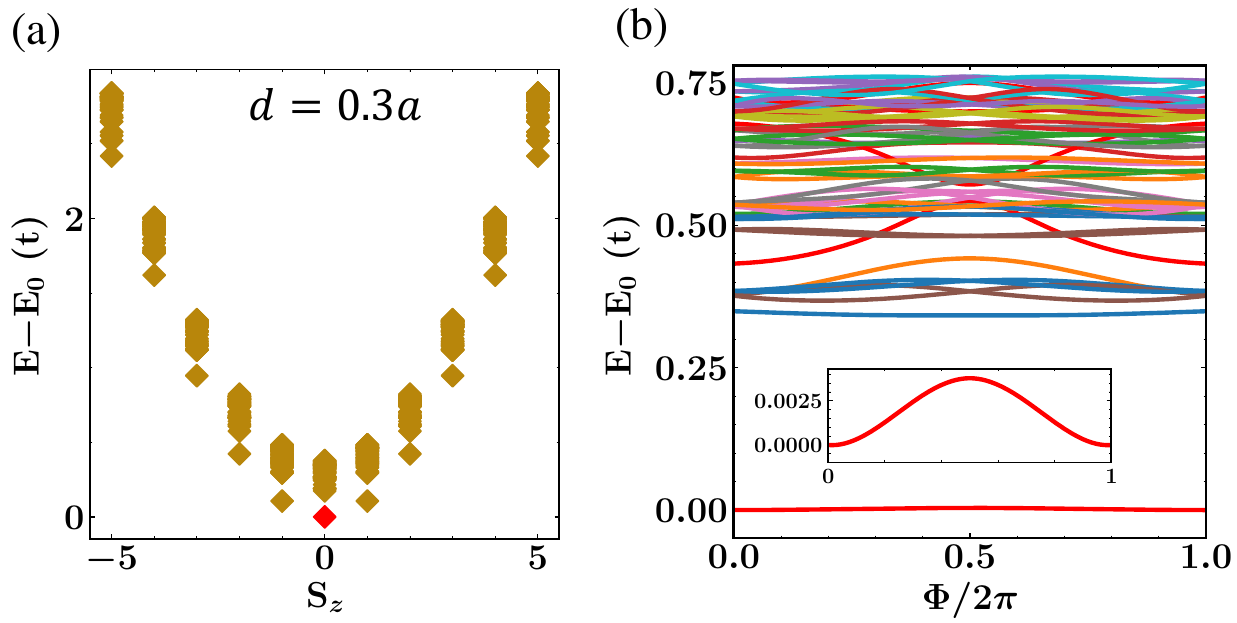}
\caption{\label{figs1} \textbf{Supplementary ED results for $v_T=1$.} (a) Energy spectra of $H$ at $v_T=1$,  $d=0.3a$, and $t_{\perp}=0$ with varying $S_z$ for $N_s=6\times4\times3$ system size. Red diamonds denote the ground states while yellow denotes the excited states. (b) Spectral flow under flux insertion in $y$-direction. The non-degenerate ground state does not mix with any excited state throughout flux insertion.}
\end{figure}

We calculate the energy spectra of $H$ at $v_T=1$, finite $d=0.3a$, and multiple $S_z$. In contrast to the case of negligible $d$ (Fig.~\ref{fig2}(c)), in this case, balanced layers with $S_z=0$ becomes energetically favorable as shown in Fig.~\ref{figs1}(a). We also calculate spectral flow of the system at $v_T=1$ under twisted boundary condition in y-direction as plotted in Fig.~\ref{figs1}(b). As is evident in the figure, after insertion of one quantum flux states get back to their initial configuration, and there is no mixing between the ground state and excited states throughout the flux insertion. Thus, the ground state has a total Chern number of 1, which also indicates a quantized Hall conductance of $\sigma_H=e^2/h$ \cite{9}.
\subsection{Halperin’s (3,3,3) state}
\begin{figure}
\centering
\includegraphics[scale=0.38]{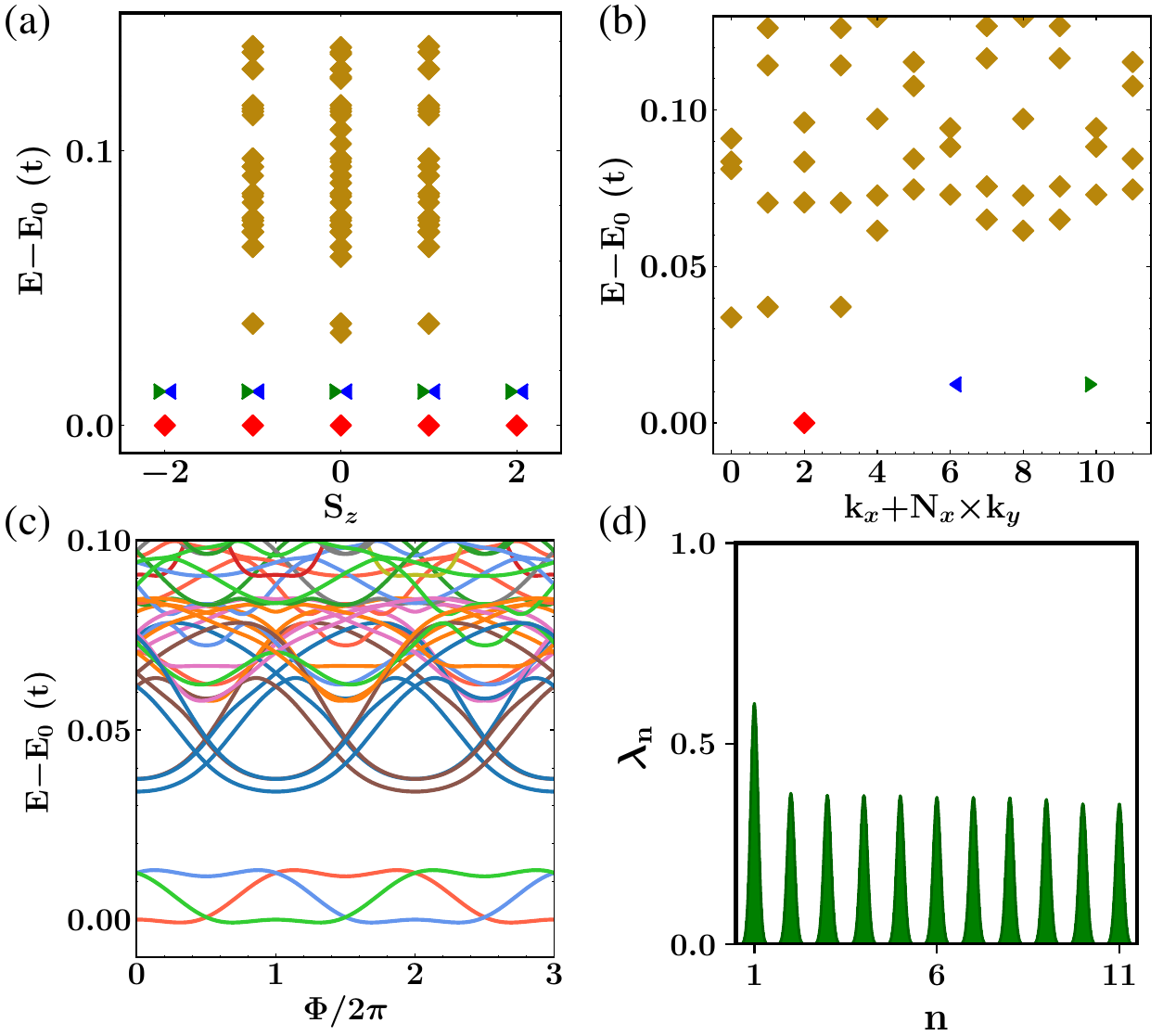}
\caption{\label{figs2} \textbf{ED calculation results at $v_T=1/3$.} (a) Energy spectra of $H$ at $d\rightarrow0$ and $t_{\perp}=0$ with varying $S_z$ for $N_s=6\times4\times3$ system size. Red diamonds, blue and green triangles denote the ground-state manifold while yellow denotes the excited states. (b) Momentum-resolved spectra of $H$ at $d=0$, $t_{\perp}=0$, and $S_z=0$. The three quasi-degenerate ground states are denoted by red, blue and green color respectively. (c) Spectral flow of the system considered in (b) under flux insertion in $y$-direction. (d) Eigenvalues of excitonic reduced two-body density matrix for (b) plotted in descending order.}
\end{figure}

In addition to the case of $v_T=1$, it is expected that similar excitonic condensate state could be present at $v_T=1/3$ described by Halperin’s (3,3,3) wavefunction \cite{2}. But so far, no experimental evidence for such a state has been reported. In the case of bilayer Kagome, we study the characteristics of system at $v_T=1/3$ with negligible $d$ and $t_{\perp}$. In Fig.~\ref{figs2}(a), one clearly sees the signature of “\textit{which layer}” uncertainty with degenerate ground-state manifold at multiple $S_z$. Different from $v_T=1$, the ground state manifold in this case has a three-fold degeneracy slightly lifted due to finite size effects. Moreover, as can be seen from Fig.~\ref{figs2}(b), the three states belong to the momentum sectors (2,0), (2,1), and (2,2) respectively, as expected from generalized Pauli’s principle [1], implying fractional statistics and non-trivial topology. In Fig.~\ref{figs2}(c) we plot the spectral flow of these states under flux insertion in $y$-direction. The three states warp around each other and go back to their original configuration after insertion of three flux quanta while the gap never closes. This correlation indicates a fractional Hall conductance \cite{9}. We also check for the exitonic off-diagonal long-range order by plotting eigenvalues of two-body reduced density matrix in Fig.~\ref{figs2}(d) and find the existence of one large eigenvalue although not as prominent as for the case of $v_T=1$. Our results here suggest that Halperin’s (3,3,3) state featuring excitonic condensate with fractional Hall conductance can in principle be realized in bilayer Kagome system by only controlling the filling factor of electrons in the system in contrast to tuning the magnetic field in BQH setup.
\begin{figure}
\centering
\includegraphics[scale=0.4]{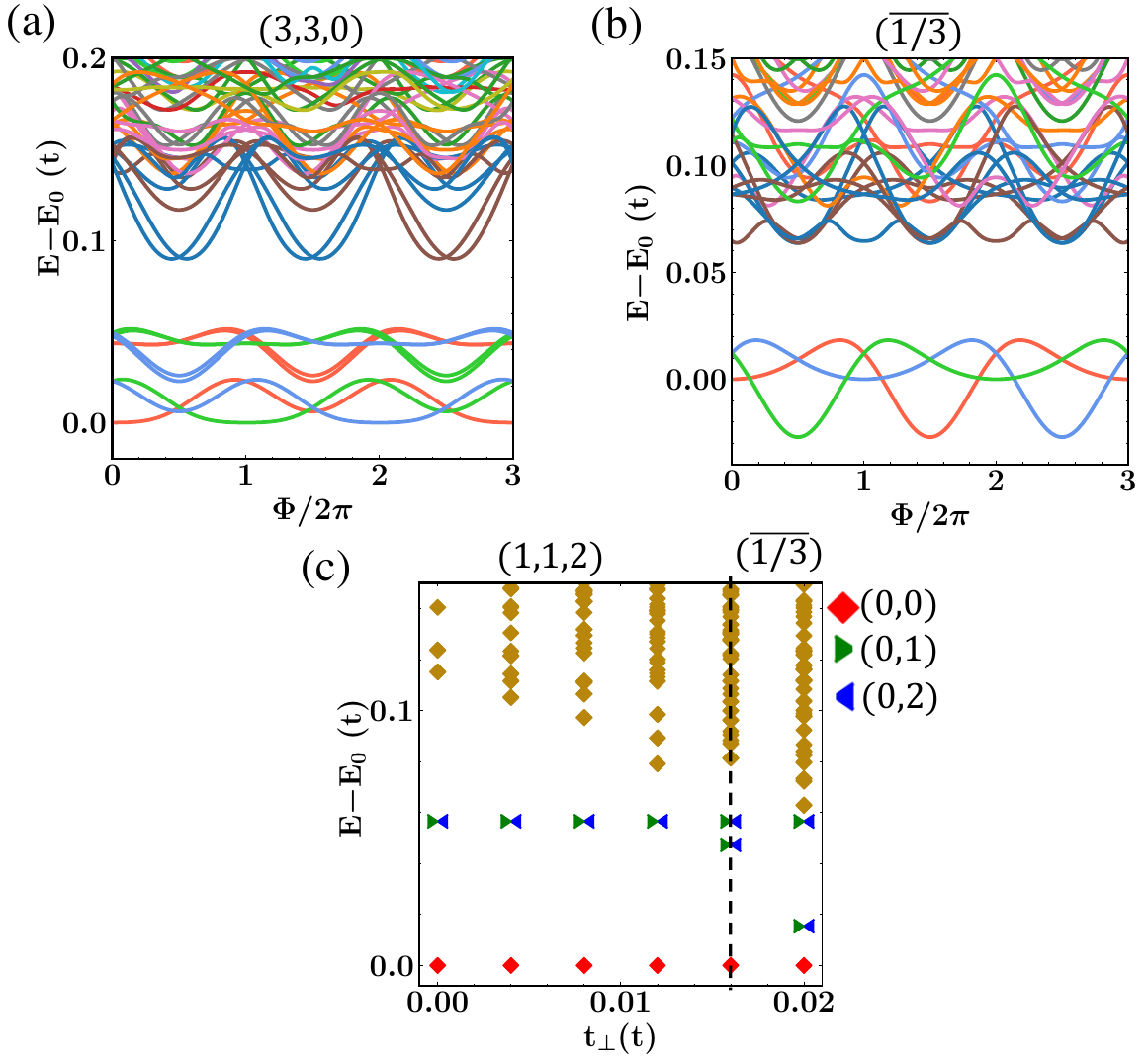}
\caption{\label{figs3} \textbf{Supplementary ED results at $v_T=2/3$.} (a) Spectral flow of the many-body states with 9-fold degenerate ground-state manifold indicating Halperin’s (3,3,0) state under flux insertion in $y$-direction calculated for $H$ at $v_T=2/3$, $t_{\perp}=0$, $d=2.25a$, and $S_z=0$. (b) (a) Spectral flow of the many-body states with 3-fold degenerate ground state manifold indicating ($\overline{1/3}$) state under flux insertion in $y$-direction calculated for $H$ at $v_T=2/3$, $t_{\perp}=0.16t$, $d=2.0a$, and $S_z=0$.  (c) Tunneling-driven phase transition from 3-fold degenerate spin-singlet (1,1,2) to 3-fold degenerate $\overline{1/3}$ state at $v_T=2/3$, $d=0.075a$, and $S_z=0$. Gap closes at $t_{\perp}=0.016t$.}
\end{figure}
\subsection{$v_T=2/3$ states}
We identified three phases at $v_T=2/3$ for bilayer Kagome lattice. For Halperin’s (3,3,0) state $K=\begin{bmatrix} m & n \\ n & m \end{bmatrix}$, $\mathrm{det}⁡K=9$, $p=2$, and $q=3$, implying $N^{\prime}=3$. For finite system size $4\times 3$, $N_e=8$, and since we are working with balanced layers ($S_z=0$), $N_i=4$, implying $N_e/(N_i p)=1$ (odd). Therefore, $N_0=N^{\prime}$, and $N_B=0$, meaning the three degenerate states corresponding to $N^{\prime}=3$ should lie at (0,0). Using Pauli’s generalized principle, there should be additional 3 states at (0,1), and (0,2) each.\par
Here we also preform the spectral-flow analysis for the Halperin’s (3,3,0) (Fig.~\ref{figs3}(a)) and particle-hole conjugate to Laughlin’s $1/3$ ($\overline{1/3}$) (Fig.~\ref{figs3}(b)) states, to provide further evidence for their topologically non-trivial characteristics.  It is only after insertion of three quantum fluxes that the states get back to their initial configuration, and there is no mixing between them and excited states throughout the flux insertion, implying a fractional Hall conductance. In Fig.~\ref{figs3}(c) we also show a tunneling-driven transition from spin-singlet (1,1,2) state to $\overline{1/3}$ state indicated by gap closing and re-opening at $t_{\perp}=0.016t$.
\begin{figure}
\centering
\includegraphics[scale=0.4]{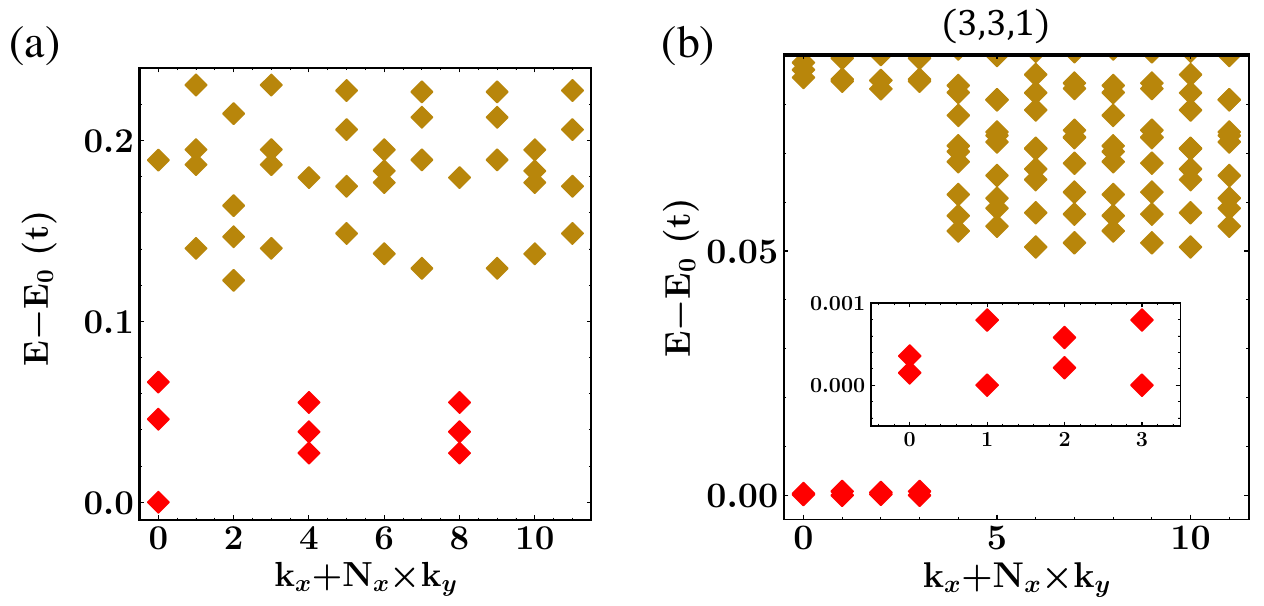}
\caption{\label{figs4} \textbf{Possible non-Abelian states in bilayer Kagome lattice.} (a) Energy spectra of $H$ at $v_T=2/3$,  $d=0.3a$, $t_{\perp}=0$, and $S_z=0$ for $N_s=6\times4\times3$ system size with attractive interlayer interaction. Red diamonds denote the 9-fold degenerate ground states while yellow denotes the excited states. (b) Energy spectra of $H$ at $v_T=1/2$, $d=0.3a$, $t_{\perp}=0$, and $S_z=0$ for $N_s=6\times4\times3$ system size. Red diamonds denote the 8-fold degenerate ground states while yellow denotes the excited states.}
\end{figure}
\subsection{Possible Abelian/non-Abelian states}
We also consider the case of $v_T=1/2$ at a finite $d=0.3a$, negligible $t_{\perp}$, and $S_z=0$. The ground state in this case is described by Halperin’s (3,3,1) wavefunction \cite{3}. Here with $q=2$, and $N^{\prime}=4$, the $K$-matrix analysis requires an 8-fold ground-state degeneracy. The four states corresponding to $N^{\prime}=4$ belong to sectors (0,0), (0,$\overline{N}/2=3$), ($\overline{N}/2=3$,$\overline{N}/2=3$), and ($\overline{N}/2=3$,0). Since $N_x=4$, and $N_y=3$, this corresponds to two states at (0,0) and (3,0) each, in addition to $q=2$ COM degeneracy. The 8-fold degenerate ground states, hence, should belong to the (0,0), (1,0), (2,0), and (3,0) momentum sectors, each two-fold degenerate. In Fig.~\ref{figs4}(a), we plot the energy spectrum for this lattice with 8-fold degenerate ground-state manifold exactly matching the $K$-matrix analysis.\par
In addition, a non-Abelian interlayer Pfaffian state with 9-fold ground degeneracy is proposed for $v_T=2/3$ BQH when interlayer interaction becomes slightly attractive \cite{4}. We see a similar 9-fold degeneracy of ground state manifold at negative interlayer interaction strength in our calculations as shown in Fig.~\ref{figs4}(b).
\subsection{Material Candidate}
\begin{figure}
\centering
\includegraphics[scale=0.35]{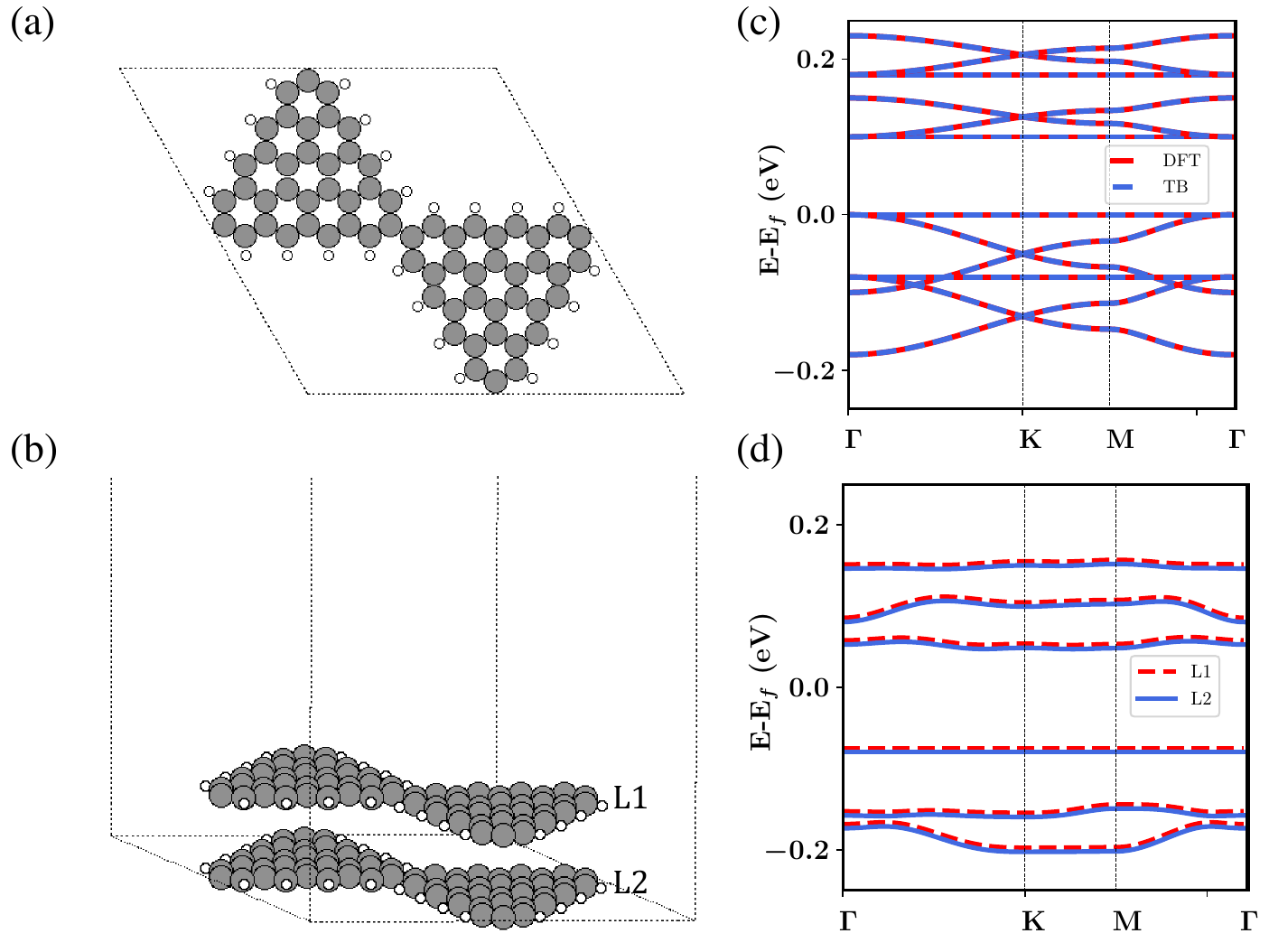}
\caption{\label{figs5} \textbf{Material candidate for realizing anomalous BQH effect.} (a) Top view of unit cell of the bilayer superatomic graphene structure used in DFT calculations. (b) Side view of (a). (c) DFT band structure fitted with TB parameters. (d) Model TB band structure using parameters $t_1=0.115 eV$ ;$t_2=0.07t_1$  ;$t_3=0.11t_1$  ;$t_5=-0.2t_1$  ; $t_{\perp}=0.0t_1$; $\lambda=0.27t_1$. Valence FBs are isolated with high flatness ratio.}
\end{figure}

We consider bilayer superatomic graphene lattice as a potential material system for realizing the exotic ABQH states without magnetic field. Single layer $9\times9$ superatomic graphene lattice has been recently proposed as a material candidate for triplet excitonic insulator \cite{5}. The band structure of monolayer has two sets of Kagome bands inverted around the Fermi-level with FBs of opposite chirality facing each other and forming conduction and valence FB \cite{5}. We consider a bilayer of this material without a tunneling barrier in between the layers (Fig.~\ref{figs5}(a),(b)) and perform first-principles calculations to calculate the band structure, as shown in Fig.~\ref{figs5}(c). These calculations were performed using the Perdew-Burke-Ernzerhof-type generalized gradient approximation for the exchange- correlation functional, as implemented in the Vienna ab-initio simulation package \cite{6}. The kinetic energy cutoff of 420 eV on a $6\times6\times1$ Monkhorst-Pack $k$-point mesh was chosen. A vacuum thickness of 44 \AA was used to avoid spurious interactions between adjacent layers. Atomic positions were fully relaxed until residual forces were less than 0.1 meV/\AA. We further fit the bands to a tight-binding (TB) bilayer model with tunneling and find the fitted TB parameters as, $t_1=0.115 eV$ ; $t_2=0.07t_1$ ; $t_3=0.11t_1$ ; $t_{\perp}=0.7t_1$, where $t_n$ is the $n^{th}$ NN hopping parameter and $t_{\perp}$ is the tunneling strength between the layers. The TB calculated band structure is shown in Fig.~\ref{figs5}(c).\par
In order to utilize this material for realizing ABQH states, one needs to introduce a tunneling barrier between the layers, so as to tune the parameter $t_{\perp}$ to a smaller value. In addition, one needs to isolate the FBs from other bands. Similar to previous work done for Kagome lattice, one can choose unusual hopping and spin-orbit coupling strengths to achieve a desired flatness ratio. Here, in addition to the above-mentioned TB parameters, we use $t_5=-0.2t_1$, and $\lambda=0.27t_1$, where $\lambda$ is the spin-orbit coupling strength with a similar form as in Eqn.~\ref{eqn1} in the main text. Such parameters can be experimentally realized using periodic modulation (Floquet engineering) as has been shown earlier \cite{7}. This procedure, hence, allows us to achieve two isolated valence bands each belonging to individual superatomic graphene layer (Fig.~\ref{figs5}(d)). Using a gate voltage one can tune the filling of these bands (Fig.~\ref{fig1}(c)) and achieve ABQH states. 

\bibliography{manuscript}

\end{document}